\documentclass{elsart}

\begin{document}

\begin{frontmatter}

\title{Neutrino oscillations and Big Bang
Nucleosynthesis\thanksref{label1}}
\thanks[label1]{Invited talk at the NUFACT'01 workshop in Tsukuba, Japan
24-30 May 2001.}

\author{Nicole F. Bell}
\ead{n.bell@physics.unimelb.edu.au}

\address{School of Physics, Research Centre for High Energy Physics\\
The University of Melbourne, Victoria 3010 Australia}

\begin{abstract}
We outline how relic neutrino asymmetries may be generated in the early
universe via active-sterile neutrino oscillations.  We discuss possible 
consequences for big bang nucleosynthesis, within the context of a 
particular 4-neutrino model.
\end{abstract}

\end{frontmatter}

\section{Introduction}

The implications of neutrino masses in cosmology and astrophysics 
are numerous.  One fascinating effect, is that of relic neutrino 
asymmetries, which may be produced 
if there is mixing between active and sterile neutrino species \cite{ftv}.

Neutrino asymmetries are interesting since they affect big bang 
nucleosynthesis (BBN), altering the primordial helium yield.  
BBN is particularly 
sensitive to an asymmetry between the $\nu_e$ and $\overline{\nu}_e$.
Additionally, since the matter potential arising from forward scattering is 
proportional to the asymmetry, a large asymmetry will suppress 
active-sterile oscillation modes which could otherwise equilibrate 
sterile neutrino species around the time of BBN.

\section{Relic neutrino asymmetries}

First consider two-flavour mixing between an active neutrino,
$\nu_{\alpha}$, and a sterile neutrino, $\nu_s$.
For each neutrino flavour we define an asymmetry
\begin{equation}
L_{\alpha}= \frac{n_{\nu_\alpha}-n_{\bar{\nu}_\alpha}}{n_\gamma},
\end{equation}
where $n_{\nu}$ and $n_{\gamma}$ are the neutrino and 
photon number densities, respectively.
The neutrino momentum distributions, $N(p)$, are such that
\begin{equation}
n=\int N(p)dp,
\end{equation}
and in thermal equilibrium are just given by Fermi-Dirac distributions
\begin{equation}
N_{{\rm eq}}(p)=
\frac{1}{2\pi^2}\frac{p^2}{1+\exp{\left(\frac{p}{T}+\tilde{\mu}\right)}}.
\label{nic}
\end{equation}
Here $T$ is the temperature and 
$\tilde{\mu} \equiv \mu/T$, with $\mu$ being the neutrino chemical 
potential.  In equilibrium, $L_{\alpha}$ is related to $\mu$ as per
\begin{equation}
L_{\alpha} \simeq -{1 \over 24\zeta(3)}
\left[ \pi^2(\tilde{\mu}_{\nu} - \tilde{\mu}_{\bar \nu})
\right. 
- \left. 6(\tilde{\mu}_{\nu}^2 - \tilde{\mu}^2_{\bar \nu}) \ln 2 
+ (\tilde{\mu}^3_{\nu} - \tilde{\mu}^3_{\bar \nu}) \right].
\end{equation}

We wish to track $L_{\alpha}$ as the universe evolves, with the period of 
interest being the epoch between $T \sim 100 {\rm MeV}$ and $T \sim 1 {\rm MeV}$. 
The evolution of the neutrino ensemble will be controlled by (i) oscillations, 
(ii) collisions of the neutrinos with other particles, resulting in decoherence of 
the oscillations, and (iii) the expansion of the universe, which redshifts energies 
and momenta.

The matter affected mixing angle, $\theta_m$, is
\begin{equation}
\sin^22\theta_m = \frac{\sin^22\theta_0}{\sin^22\theta_0+
(b \pm a -\cos2\theta_0)^2},
\end{equation}
where $\theta_0$ is the mixing angle in vacuum.
The term $(b \pm a)$ arises from the effective matter potential 
$V=(b\pm a)\delta m^2/2p$, where -/+ corresponds to $\nu/\bar{\nu}$
and $\delta m^2$ is the mass 
squared difference of the two neutrino mass eigenstates.
The functions $a$ and $b$ are given by
\begin{eqnarray}
\label{eq:a}
a(p) &=&-\frac{4\sqrt{2}\zeta(3)G_FT^3L^{(\alpha)}p}
{\pi^2\delta m^2}, \nonumber \\
b(p) &=&-\frac{4\sqrt{2}\zeta(3)G_FT^4A_{\alpha}p^2}
{\pi^2\delta m^2 M_W^2},
\end{eqnarray}
where $A_e (A_{\tau,\mu})\simeq 17 (4.9)$ \cite{rn}.
The function $L^{(\alpha)}$ is defined as
\begin{equation}
L^{(\alpha)} = L_{\alpha} + L_{\tau} + L_{\mu} 
+ L_{e} + \eta,
\end{equation}
where $\eta \sim10^{-10}$ is due to the baryon asymmetry.
Note that the mixing angle is dependent on the size of the asymmetry, 
which makes the evolution of the asymmetry non-linear.

There are two distinct phases in the growth of a neutrino asymmetry.
The first is a collision dominated stage, which occurs at high temperature.     
The total collision rate of a $\nu_e$ ($\nu_{\mu}$ or $\nu_{\tau}$) of momentum 
$p$ with particles in the background plasma is approximately
\begin{equation}
\Gamma (p) \simeq 4.0 (2.9) \frac{p}{3.15 T} T^5.
\end{equation}
These collisions tend to destroy the coherent superposition of flavour states.
In this regime, it can be shown that the growth of the neutrino asymmetry is 
governed by the equation
\begin{eqnarray}
\label{hightemp}
\frac{d L_{\alpha}}{dt}  \simeq  \frac{\pi^2}{2\zeta(3)T^3} 
\int dp  \left( N^+_{\alpha}-N^+_s \right) 
\frac{s^2 \Gamma a(c-b)}{[x+(c-b+a)^2][x+(c-b-a)^2]},
\end{eqnarray}
where $c \equiv \cos 2\theta_0$, $s \equiv \sin 2\theta_0$,
\begin{equation}
x \equiv \left[ \frac{p \Gamma (p)}{\Delta m^2} \right]^2, \;\;\;
N^{+} = \frac{1}{2}(N_{\nu}+ N_{\overline{\nu}}).
\end{equation}
Eq.(\ref{hightemp}) can be derived either from a heuristic picture \cite{ftv} that 
views collisions as measurement-like interactions, or can be extracted from the 
exact Quantum Kinetic Equations \cite{bvw}.  It is valid when the collision rate 
is large and the evolution is adiabatic - see \cite{bvw} for details. 

Note that $dL_{\alpha}/dt \propto a \propto L^{(\alpha)}$, which can lead 
to runaway positive feedback and thus rapid growth of $L_{\alpha}$.  
The mixing parameters
for which this occurs are specified by
$|\delta m^2| \stackrel{>}{\sim} 10^{-4}\rm{eV}^2$ and
$\sin^2 2\theta_0 \stackrel{>}{\sim}  10^{-10}$, and
we require lighter of the 
two mass eigenstates to approximately coincide with the $\nu_s$. 
Typically, the asymmetry will reach values 
$L_{\alpha} \sim O(10^{-5})$ in this phase.
Note that the sign of the asymmetry, which depends on the initial conditions, 
can be ambiguous \cite{dibarifoot}.

The second phase is that lower temperature.  Here the collisions are less 
important, and coherent MSW transitions drive the asymmetry to values 
$L_{\alpha} \sim O(0.1)$ \cite{fv}.\footnote{Ref.\cite{dhps} disputed the 
production of large 
asymmetries. However, this issue has been resolved, with \cite{dhps} shown 
to be in error due to the use of an approximation outside its region of 
validity \cite{comment}.}  If, for example, the asymmetry is positive, the 
neutrino MSW resonance occurs at a much higher momentum value than that of the 
antineutrinos and may be neglected.
The rate at which the asymmetry grows is then
\begin{equation}
\label{res}
\frac{dL_{\nu_{\alpha}}}{dT} = 
-X\left|\frac{d}{dT}\left(\frac{p_{{\rm res}}}{T}\right)\right|,
\end{equation}
where $X$ is the difference between 
the number of $\bar{\nu}_{\alpha}$ and
$\bar{\nu}_s$ at the resonance.
Eq.(\ref{res}) quantifies the rate at which the resonance moves through the 
momentum distribution, and holds when transitions are adiabatic and the 
resonance width is small.

\section{Big bang nucleosynthesis and a four neutrino model}

We wish to determine the impact of such asymmetry production on the BBN 
light element abundances, and in particular, on the helium yield.  
In general, the results are somewhat model 
dependent (compare, say, \cite{fv} with \cite{bfv}.)  

The results are particularly sensitive to the size of the $\nu_e$ asymmetry, 
which directly affects the neutron/proton ratio through the reaction rates
\begin{eqnarray}
\label{eq:rates}
\lambda(n\rightarrow p)\simeq \lambda(n \nu_e \rightarrow p  e^-) 
+ \lambda(n e^+ \rightarrow p  \bar{\nu}_e), \nonumber \\
\lambda(p\rightarrow n) \simeq \lambda(p  e^- \rightarrow n  \nu_e) 
+ \lambda(p  \bar{\nu}_e \rightarrow n  e^+).
\end{eqnarray}
The n/p ratio determines the helium mass fraction $Y_P$ via the equation
\begin{equation}
\frac{dY_P}{dt}=-\lambda(n\rightarrow p)Y_P+\lambda(p\rightarrow n)(2-Y_P).
\end{equation}

For typical values of $\delta m^2$, one must compute the growth of the 
asymmetry and the effect on BBN simultaneously, as the asymmetries will 
not have reached their final values before the BBN epoch commences.

Momentum modes which are depleted due to the oscillations, are refilled by 
scattering and annihilation processes as per
\begin{equation}
\frac{d}{dt}\left( \frac{N_{{\rm actual}}}{N_0} \right) = \Gamma(p) 
\left( \frac{N_{{\rm eq}}}{N_0}-\frac{N_{{\rm actual}}}{N_0} \right), 
\label{eq:refill}
\end{equation}
where $N_{{\rm eq}}$ is the equilibrium distribution in Eq.(\ref{nic}), and 
$N_0 \equiv \left. N_{{\rm eq}}\right|_{\tilde{\mu}=0}$. 

The example we consider is a particular ``2+2'' style 4-neutrino model, 
consisting of a maximally mixed $\nu_{\mu} - \nu_{\tau}$ pair, separated 
by a mass gap from $\nu_e - \nu_s$.
The oscillations modes most important in the evolution of the various flavour 
neutrino asymmetries will be the ones with the largest $\delta m^2$.  
Specifically, the $\nu_{\mu,\tau}-\nu_s$ oscillation modes will create 
large $L_{\mu}$ and $L_{\tau}$ asymmetries, and the  $\nu_{\mu,\tau}-\nu_e$
oscillation modes will transfer a small amount of this asymmetry to the 
electron neutrinos.  See \cite{bfv} for further details.

Numerically integrating the appropriate generalisations of eq.(\ref{res}),
we find that the magnitude of the final lepton asymmetries generated are
\begin{eqnarray}
|L_{\mu}|/h = |L_{\tau}|/h &\simeq& 0.16, \nonumber\\
|L_e|/h &\simeq& 6.7\times10^{-3},
\end{eqnarray}
for  $\delta m^2 \stackrel{>}{\sim} 4 {\rm eV} $, where 
$h=T_{\nu}^3/T^3_{\gamma}$.

Integrating the BBN reaction rates, eq. (\ref{eq:rates}), the direct 
effect of the $\nu_e$ asymmetry on the $n/p$ ratio, and hence 
$Y_P$, is determined.  
We obtain $\delta Y_P\simeq \pm 0.0023$, corresponding to an effective change 
in the number of neutrinos of $\delta N_{\nu}^{eff}\simeq \pm 0.19$. 
In addition, there is a small direct change to the energy density, leading overall to
\begin{eqnarray}
\delta N_{\nu}^{{\rm eff}}\simeq-0.3,  \;\; {\rm positive \;\; asymmetry}, \nonumber \\
\delta N_{\nu}^{{\rm eff}}\simeq+0.1,   \;\; {\rm negative \;\; asymmetry},
\end{eqnarray}
for $\delta m^2 \stackrel{>}{\sim} 3 {\rm eV}^2$.

The value of $N_{\nu}^{{\rm eff}}$ obtained in this way must then be compared 
with that calculated from the observed light abundances \cite{olive}.  It is not 
yet clear whether the observed helium and deuterium abundances can be accommodated 
within the standard BBN scenario with $N_{\nu}^{{\rm eff}}=3$, and it is possible 
that a small $\nu_e$ asymmetry such as that discussed here would help to provide a 
better fit \cite{hansen}. 
Recent measurement of the cosmic microwave background radiation (CMBR) anisotropy 
also provides a determination of the number of neutrinos in the early 
universe \cite{hannestad}, although the constraints are, as yet, not as strong 
as those arising from 
BBN.\footnote{Note, however, that the effective number of neutrinos extracted from 
the CMBR and BBN data are not necessarily the same parameter.  The two determinations 
are complementary, since the BBN quantity depends on flavour through its sensitivity 
to the $\nu_e$ asymmetry, while the CMBR measures the total energy density residing 
in neutrinos (at a much later epoch of the universe.)} However, future measurements 
are expected to pin down the effective neutrino number very precisely 
\cite{hannestadraffelt} and thus will be of great use in constraining models which 
feature the mixing of active and sterile neutrinos.

\end{document}